\documentclass[aps,12pt,preprint,floatfix,tightenlines,showkeys,showpacs]{revtex4}
\usepackage{epsfig}
\usepackage{graphicx}
\usepackage{amssymb}
\newcommand{\nn}{\nonumber\\&&}

\newcommand{\pslash}{\not\hspace{-0.7mm}p}
\newcommand{\kslash}{\not\hspace{-0.7mm}k}
\newcommand{\ben}{\begin{displaymath}}
\newcommand{\een}{\end{displaymath}}
\newcommand{\be}{\begin{equation}}
\newcommand{\ee}{\end{equation}}
\newcommand{\bea}{\begin{eqnarray}}
\newcommand{\eea}{\end{eqnarray}}

\newcommand{\eq}[1]{Eq.~(\ref{#1})}

\newcommand{\bfp}{{\bf p}}

\begin{document}

\title{\bf 
{Taming the Pion Cloud of the Nucleon}}

\author{  Mary Alberg$^{1,2}$,  Gerald A. Miller$^2$}

\affiliation{$^1$Department of Physics, Seattle University, Seattle, WA 98122, USA }

\affiliation{$^2$Department of Physics,
University of Washington, Seattle, WA 98195-1560}

\date{\today}

\begin{abstract}

{We present a light-front determination of the pionic contribution to the nucleon self-energy, $\Sigma_\pi$, to second-order in  pion-baryon coupling constants that allows the pion-nucleon vertex function to be treated in a  model-independent manner  constrained by experiment. The pion mass $\mu$  dependence of  $\Sigma_\pi$ is  consistent with chiral perturbation theory results for small values of $\mu$ and  is  also   linearly dependent on  $\mu$ for larger values, in accord with the results of  lattice
QCD calculations.  The  derivative of $\Sigma_\pi$ with respect to $\mu^2$ yields the dominant contribution to the pion content, which is consistent with the $\bar{d}-\bar{u}$ difference observed experimentally  in the violation of the Gottfried sum rule. }
\end{abstract}\pacs{13.75.Gx, 12.39.Fe, 12.38.Gc,  13.60.Hb   }
\keywords{meson loop, nucleon self-energy, deep inelastic scattering}
\preprint{NT@UW-12-01}

\maketitle
\begin{figure}
\includegraphics[width=10.991cm,height=10.5cm]{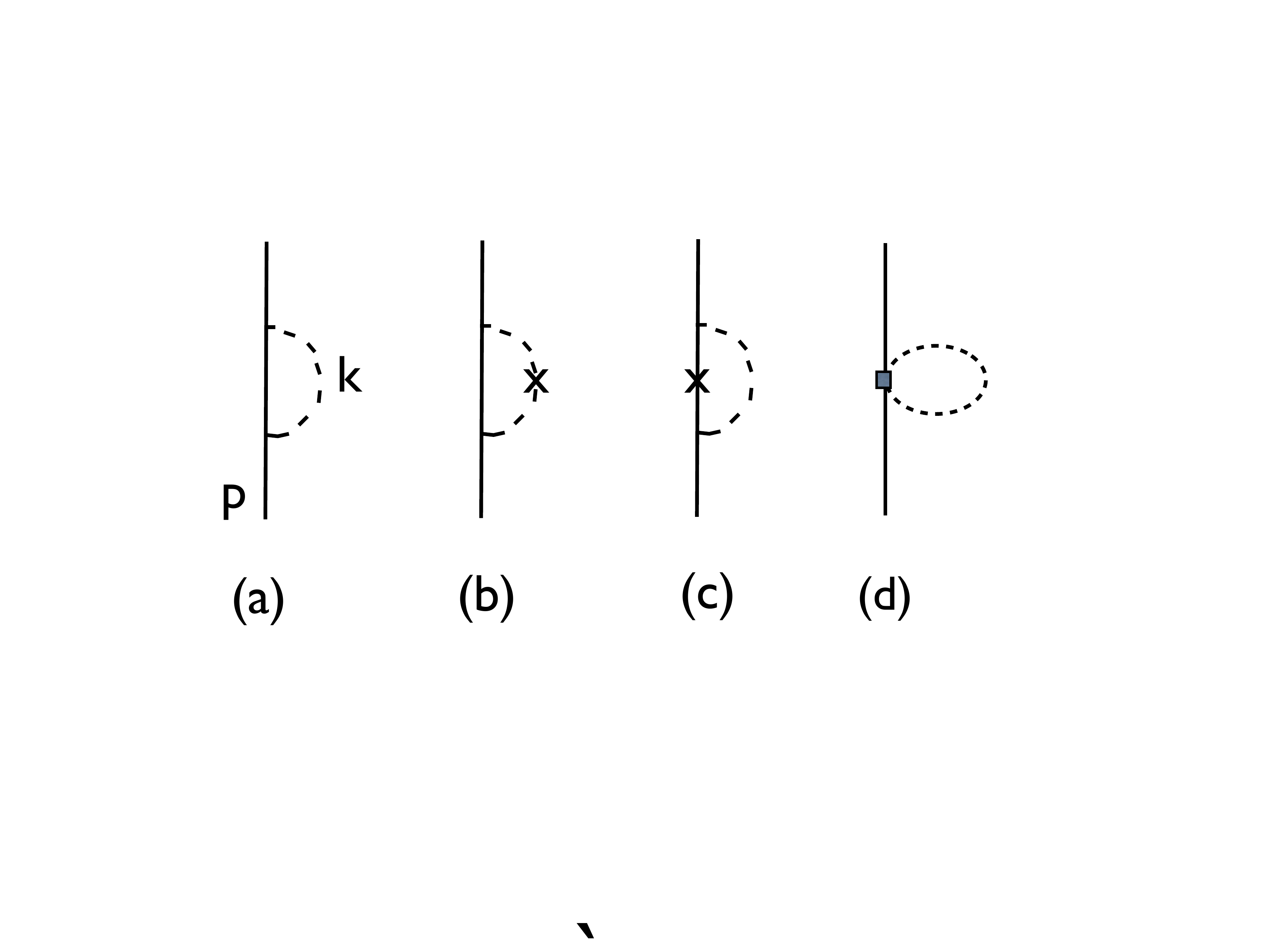}
\caption{(a) Pionic (dashed line) contribution to the nucleon (solid line) self-energy. (b) External interaction, x, with the pion (c) External interaction, x, with the intermediate nucleon. (d) Effect of 2$\pi$-nucleon interaction.}\label{fig:diag}\end{figure}

     
\noindent
 Understanding the pion and its interaction with and amongst nucleons is a necessary step in 
 learning how QCD describes the interaction and existence of atomic nuclei.
 As a  nearly massless excitation of the QCD vacuum with pseudoscalar quantum numbers, the 
pion plays a central role in particle and nuclear physics as a harbinger of spontaneous symmetry breaking. 
The pion is associated with large distance structure of the nucleon~\cite{Theberge:1980ye,Strikman:2009bd} and the longest ranged component of the nucleon-nucleon force~\cite{Machleidt:1989tm}. 
In  lattice QCD calculations  the nucleon mass depends on an input value of the  quark mass, which generates  a pion mass $\mu$, and extrapolation formulae depending on $\mu$ are typically used~\cite{ Aoki:1999yr,Leinweber:1999ig,Beane:2004ks,Young:2009zb}
 (see the review~\cite{hep-lat/0608010}.)  In addition, 
 the pion  cloud plays an important role in deep inelastic scattering on the nucleon, especially in understanding the violation of the Gottfried sum rule~\cite{Thomas:1983fh,Henley:1990kw}.  
  \vskip0.2cm\noindent 
Phenomenological calculations  of pion-nucleon interactions are beset with uncertainties related to the dependence of the vertex function on momentum transfer and on the possible dependence upon the virtuality (difference between the square of the four-momentum and mass squared) of any intermediate nucleon or baryon. 
Moreover, modern treatments    of spin 3/2 baryons such as the $\Delta$ (baryon excitation of lowest mass) within the Rarita-Schwinger (RS)~\cite{Rarita:1941mf} formalism have  been problematic as discussed in \cite{nucl-th/9905065}. The pathologies of the $\pi N\Delta$ coupling have long been known~\cite{Nath:1971wp,Nath:1979wr,Benmerrouche:1989uc, Pascalutsa:1998pw,Davidson:1991xz}.
The aim of the present letter is to develop and apply a method that is free of  those ambiguities.
 \vskip0.2cm\noindent 
 As a  specific example, consider the role of the pion cloud in 
deep inelastic scattering. This is related to the pion contribution to the nucleon self-energy of    
Fig.~\ref{fig:diag}a.  One needs to include   the term 
in which the virtual photon interacts with the pion~\cite{Sullivan:1971kd}, Fig.~\ref{fig:diag}b, but one also needs to include
the effects of the virtual photon hitting the nucleon, 
 Fig.~\ref{fig:diag}c.
 Conservation of momentum and  charge would seem to require that the argument of the vertex function
 depends on the square of the invariant mass of the intermediate pion-baryon system ($s$)~\cite{Holtmann:1996be}.
Taking the form factor to have the standard form of depending on the square of the four-momentum transfer, between the initial nucleon and intermediate baryon ($t$), while natural, popular and effective \cite{Koepf:1995yh},\cite{Strikman:2009bd} seemingly 
disagrees with charge and momentum conservation according to ~\cite{Holtmann:1996be}.

But chiral symmetry  (limit of vanishing pion mass) provides strong guidance.
It is known that  the $\pi N$ vertex function $ G_{\pi N}(t)$  and the nucleon axial form factor are related  by the 
generalized Goldberger-Treiman relation~\cite{Thomas:2001kw}):
 \bea MG_A(t)=f_\pi G_{\pi N}(t),\label{relate}\eea where $t$ is the square of the four-momentum transfered to the 
 nucleons, $G_A(t)$ is the axial vector form factor and  $f_\pi$ is the pion decay constant.  The result \eq{relate},
 obtained from  a matrix element of the axial vector current between two on-mass-shell nucleons,  follows from PCAC and the pion pole dominance of the pseudoscalar current. 
Using  \eq{relate} has obvious practical value because it relates an essentially unmeasurable quantity $G_{\pi N}$ with one $G_A$ that is constrained by experiments. However the $t$ dependence inherent in  \eq{relate} would seem to violate the purported consequence of momentum conservation. 
Similarly the pionic coupling between nucleons and $\Delta$ particles has an off-diagonal Goldberger-Treiman
 relation~\cite{odgt1,odgt2,arXiv:0706.3011}, obtained using similar logic:
 \bea 2MC_5^A(t)=f_\pi G_{\pi N\Delta}(t),\label{relate1}
\eea
where $C_5^A$ is the Adler form factor~\cite{Adler:1968tw,Llewellyn Smith:1971zm}, accessible in neutrino-nucleon interactions. 

The present manuscript  develops   a method that  satisfies momentum conservation, utilizes \eq{relate}  and involves only on-mass-shell nucleons. The key to removing ambiguities  lies in evaluating the relevant Feynman diagrams by carrying out the integration  over the  four-momentum $k$ by first integrating over $k^-$ (the light front energy) in such a way that the intermediate baryon is  projected onto its mass shell. This allows the use of the on-mass shell form factors Eqs.~(\ref{relate},\ref{relate1})  and is manifestly 
consistent with charge and momentum conservation.

Consider the contribution to the nucleon self-energy ${\Sigma}_\pi(N)$, involving an intermediate nucleon,  Fig.~(\ref{fig:diag}a),  given by 
 Feynman rules as 
\bea { \Sigma} _\pi(N)=-i3 g_{\pi N}^2\bar{u}(P)\int{d^4k\over ( 2\pi)^4} {\gamma^5 (\pslash-\kslash +M)\gamma^5\over (k^2-\mu^2+i\epsilon)((p-k)^2-M^2+i\epsilon)} u(P)F^2(k^2),
\label{sigma1}\eea
where $M,\mu $ are the  nucleon and  pion masses. 
The quantity $P$ represents the
  nucleon momentum and spin, $(p,s)$, evaluated in the proton rest frame.  We use the notation:
 $G_{\pi N}(t)\equiv g_{\pi N} F(t)={M\over f_\pi}G_A(t),$ with  $G_A(0)=1.267\pm0.04,\;M=0.939\; {\rm GeV},f_\pi=92.6\;{\rm MeV},g_{\pi N}\equiv G_{\pi N}(0)=13.2$
with $F(0)=1$.
The term $F(k^2)$ represents the pion nucleon form factor. Its dependence on a single variable is justified only if the
pionic vertex function appears between two on-mass-shell nucleons.
In that case, one  may use  
 a dispersion relation:
\bea F(k^2)={1\over \pi}\int_{(3m_\pi)^2}^\infty dt' \;Im[F(t')]/(k^2-t').\label{disp}\eea

Performing  the spin average of \eq{sigma1}
leads to the result
\bea &&{\Sigma}_\pi(N)
={3 g_{\pi N}^2\over M } 
\int{d^4k\;F^2(k^2)\over i ( 2\pi)^4} {k\cdot p\over (k^2-\mu^2+i\epsilon)((p-k)^2-M^2+i\epsilon)} .\eea
We evaluate ${\Sigma}_\pi(N) $ using 
 light-front coordinates:
$k^\pm\equiv k^0\pm k^3,\;k^2=k^+k^--k_\perp^2$. Thus  
${\Sigma}_\pi(N)
={3 g_{\pi N}^2\over M } \int dk^+d^2k_\perp J,$ with 
\bea && J\ 
={1\over i(2\pi)^4}{1\over2} \int dk^- F^2(k^2) {k\cdot p\over  k^+(p-k)^+(k^- -{k_\perp^2+\mu^2-i\epsilon\over k^+})((p-k)^--{k_\perp^2+M^2-i\epsilon\over p^+-k^+})}.
\eea
The  expression \eq{disp} for $F(k^2)$ is not written explicitly  here because the analytic structure is the same as that of $1/(k^2-\mu^2 +i\epsilon)$.
If $0<k^+<p^+$, the first pole in $k^-$ is in the lower half $k^-$ plane (LHP) (as are the ones arising from $F(k^2)$) and the intermediate nucleon pole  is in the upper half plane (UHP). We integrate over the UHP, so that the {\it only} pole we need to consider is the one in which the intermediate nucleon is on its mass shell and the momentum $k$ is space-like. 
For $k^+<0$ and $k^+>p^+$ all of the  poles are on the same side of the real axis, and one obtains  0. We take the residue of the  integral for which  the  nucleon is on shell so that 
$k^-=p^--{M^2+k_\perp^2\over p^+-k^+}$. 
Using the residue theorem and integrating over $k^+$ leads to the result
\bea {\Sigma}_\pi(N)=- {3 g_{\pi N}^2}{\pi\over 8M(2\pi)^3}\int_0^\infty dt\; {t\;F^2(-t)\over (t+\mu^2)}\left(-{t\over M^2}+\sqrt{{t^2\over M^4}+{4t\over M^2}}\right).
\label{fn}
\eea
This result is obtained by using the pseuodscalar form of $\pi$N coupling in \eq{sigma1}, but the use of pseudovector coupling would give the same result because the intermediate nucleon is on its mass shell.

To proceed we  use a specific form of the form factor $F$,
  the commonly used dipole parametrization 
 \bea F(Q^2)=1/(1+(Q^2/M_A^2))^2,\eea
  with $M_A$ as the so-called axial mass. 
  The values of $M_A$ are given by $M_A=1.03\pm 0.04$ GeV as reviewed in \cite{Thomas:2001kw}. This range is consistent with the one reported in a later review~\cite{Bernard:2001rs}. 
   A somewhat lower value (0.85 Gev) is obtained~\cite{Bhattacharya:2011ah} if one restricts the extraction region to very low values of $Q^2$, but we need higher values to evaluate \eq{fn}.
Using this dipole parameterized  form factor $F$ gives
\bea &&{\Sigma}_\pi(N)=- {3 M g_{\pi N}^2}{\pi\over 4(2\pi)^3}\frac{1}{6 \left(\frac{4}{b}-1\right)^{5/2} (a-b)^4}\times
\nonumber\\&&
[\sqrt{(4-b) b} \left((a-b)^2 (a (b-10)+2 (b-1) b)-3 a^2 (b-4)^2 b \log
   \left(\frac{b}{a}\right)\right)\nonumber\\&&+6 \left(4 a^3+a^2 (b-6) b ((b-4) b+6)-2 a b^2
   ((b-10) b+18)-2 (b-2) b^3\right) \tan ^{-1}\left(\sqrt{\frac{4}{b}-1}\right)\nonumber\\&&+6 a b
   (b-4)^2 \sqrt{(a-4) a (b-4) b} \tan ^{-1}\left(\sqrt{\frac{4}{a}-1}\right)],\;a\equiv\mu^2/M^2,\;b\equiv M_A^2/M^2.\label{nf}\eea
To relate to chiral perturbation theory we expand in powers of $a$ up to order $\mu^4$ and $b$ around unity to obtain a very accurate  representation of the exact expression for $0\le  a\le 0.04,\;0.6\le b\le1.6$.
We find
\bea &&\widetilde{\Sigma}_\pi(N)=- {3 M g_{\pi N}^2}{\pi\over 4(2\pi)^3}[\frac{2 \pi }{27 \sqrt{3}}+\left(-\frac{1}{6}-\frac{10
   \pi }{27 \sqrt{3}}\right) a+\pi a^{3/2}+\nn\left(\left(\frac{2}{3}+\frac{104 \pi }{81 \sqrt{3}}\right) a^2-\frac{16
   \pi  a}{81 \sqrt{3}}+\frac{8 \pi }{81 \sqrt{3}}\right) (b-1)+ a^2 \left(\frac{\log
   (a)}{2}-\frac{67 \pi }{27 \sqrt{3}}-\frac{4}{3}\right)],\label{nt}\eea
   where the tilde indicates that a chiral expansion has been made.
 The term independent of the pion mass provides a  -0.222 $M$  correction to the bare nucleon mass, in contrast with an early  approach (not using the heavy baryon expansion)  which gives a contribution of formal order $M (M/4\pi f_\pi)^2$~\cite{gasser}.     
 The term of order $\mu^3$ reproduces the standard expression:
 $-3g_A^2/(32\pi f_\pi^2)\mu^3$~\cite{Langacker:1974bc}.

The next step is to include terms with an intermediate  $\Delta$, the baryon excited state of lowest mass,  which   couples strongly to the $\pi N$ system.
The effects of other intermediate baryons are not included in this first evaluation, but our technique can be applied to those states.
We use the isospin-invariant  interaction Lagrangian of  the form ${\cal L}_{\pi N\Delta}={g_{\pi N\Delta}\over 2M} \bar{\Delta}^i_\mu(\bfp')g^{\mu\nu}u(\bfp)\partial_\nu \pi^i$ +H.c.~\cite{odgt1,odgt2}   which yields the same result as the gauge invariant coupling of~\cite{nucl-th/9905065}
for an on-shell intermediate $\Delta$.  We note that $\Delta^i$ is a vector spinor in both spin and  isospin space and $ g_{\pi N\Delta}=\sqrt{6}/2G_{\pi N\Delta}(0)$, a notational relation between re-normalized coupling constants~
\cite{odgt1}.
The  contribution of the intermediate $\Delta$ to the nucleon self-energy is given by 
\bea &&
{ \Sigma} _\pi(\Delta)=i 2 ({g_{\pi N\Delta}\over 2M})^2\bar{u}(P)\int{d^4k\over ( 2\pi)^4} { (\pslash-\kslash +M_\Delta)\over (k^2-\mu^2+i\epsilon)((p-k)^2-M_\Delta^2+i\epsilon)}{(p-k)^2\over M_\Delta^2}\nn
\times P_{\mu\nu}^{(3/2)}(p-k)k^\mu k^\nu u(P)F_\Delta^2(k^2),
 \label{sigmad1}\eea
where the factor of 2 arises from the  isospin matrix element, $M_\Delta$ is the mass of the $\Delta$ and our notation for the projection operator  $P_{\mu\nu}^{(3/2)}$ is given in \cite{nucl-th/9905065}. We take the ratio of coupling constants 
to be $ ({g_{\pi N\Delta}\over g_{\pi N}})^2
= 72/25$, which is the $SU(6)$ quark model result. The form factor $F_\Delta$ is defined via  $G_{\pi N\Delta}(t)\equiv g_{\pi N\Delta} F_\Delta(t)={2M\over f_\pi}C_5^A(t)$.
Performing  the spin average 
leads to the result
\bea &&{\Sigma}_\pi(\Delta)=2({g_{\pi N\Delta}\over 2M})^2 
{1\over M}\int{d^4k\;F_\Delta^2(k^2)\over i ( 2\pi)^4} (M^2-p\cdot k+MM_\Delta){(p-k)^2\over M_\Delta^2}\nn \times{2\over3}[k^2- {\left(k\cdot(p-k)\right)^2\over  (p-k)^2}]
 {1\over (k^2-\mu^2+i\epsilon)((p-k)^2-M_\Delta^2+i\epsilon)} .\nn
\eea

We evaluate ${\Sigma}_\pi(\Delta) $ using  light-front coordinates in a procedure analogous to that used for $\Sigma_\pi(N)$.
The integral over $k^-$ is done in the upper half $k^-$ plane (UHP), so that the only pole is the one in which the intermediate $\Delta$ is on its mass shell and the momentum $k$ is space-like. The result is  
\bea&&{\Sigma}_\pi(\Delta)=-2 ({ g_{\pi N\Delta}\over 2M})^2{\pi\over M(2\pi)^3} { 1\over3}\int_0^\infty  dt\; {\;F_\Delta^2(-t)\over (t+\mu^2)}
\left (t+{1\over4M_\Delta^2}(M^2-M_\Delta^2+t)^2\right)\nn\times
 {1\over2}\left((M+M_\Delta)^2 +t\right)
\left({-t+M^2-M_\Delta^2\over 2M^2}+{1\over2M^2}\sqrt{{(M_\Delta^2-M^2+t)^2+4tM^2}}\right) .\label{deltf}
\eea

We turn to numerical evaluations.
Lattice calculations \cite{arXiv:0706.3011} indicate that the ratio
 ${G_{\pi N\Delta}(t)/G_{\pi N}(t)}$ is  constant as a function of the space-like values of $t$, thus here we use $F_\Delta(t)=F(t)$.
The integration of \eq{deltf}  yields a lengthy  closed form expression. 
To gain insight, and compare with the general form of the chiral expansion of baryon masses in QCD {\it e.g.}~ \cite{Jenkins:1990jv,Jenkins:1991ts,Bernard:1993nj}
we take $M_A=M,b=1$ and expand in $\mu/M$, $(\xi-0.72),\;\xi\equiv {M_\Delta^2-M^2\over M^2}$ to  find
\bea &&\widetilde{\Sigma}_\pi(\Delta)=-2 M({ g_{\pi N\Delta}\over 2})^2{\pi\over  (2\pi)^3} { 1\over3}[ f_1(a)+(\xi -0.72) f_2(a)] \label{dt}   \\&&
 f_1(a)\equiv-0.888 a^2+1.01 a^2 \log (a)-1.55  a^2 (\log (a)+1.20)\nn-0.402  a^2 (\log (a)+1.24)-0.00369 a+0.280 a \log
   (a)+0.310\\&&
   f_2(a)\equiv (5.48 a^2+1.46 a^2 \log (a)+2.39 a^2 (\log (a)+1.20)\nn+0.128 a^2 (\log (a)+1.24)+1.02  a+0.318 a \log
   (a)-0.0196  ),
\eea
    where the tilde indicates that a chiral expansion has been made.
The terms of order $\mu^4 \log \mu^2$ emphasized by~\cite{Leinweber:1999ig},\cite{Armour:2008ke}  are included, but  the expression also
contains  previously noted  
\cite{Bernard:1993nj}
dominating non-analytic terms of the 
form $\mu^2 \log\mu^2$.

\begin{figure}
\includegraphics[width=9.5991cm,height=8cm]{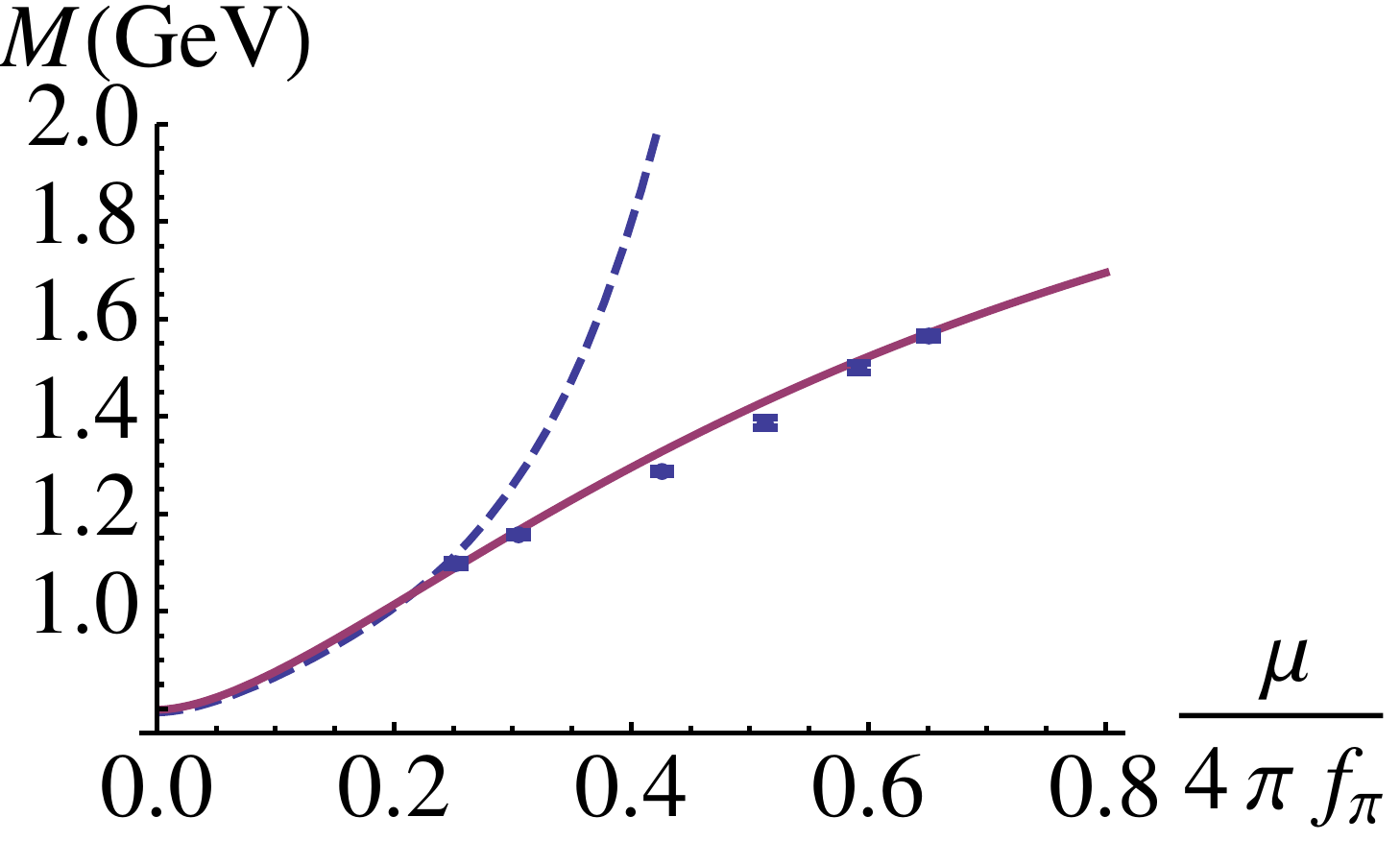}
\caption{Nucleon mass as a function of $\mu$. Square blocks: LHP lattice data~\cite{arXiv:0806.4549}. 
Solid: $\Sigma_\pi$ of \eq{tot}, \eq{nf} and \eq{deltf}. Dashed: chiral approximation $\widetilde{\Sigma}_\pi=\widetilde{\Sigma}_\pi(N)+\widetilde{\Sigma}_\pi(\Delta)$, \eq{nt} and \eq{dt}. }\label{fig:extrap}\end{figure}

The total pionic contribution to the nucleon mass $\Sigma_\pi$ is given by 
\bea \Sigma_\pi\equiv\Sigma_\pi(N)+\Sigma_\pi(\Delta),\label{tot}\eea
and the chiral approximation $\widetilde{\Sigma}_\pi$ is given by 
$ \widetilde{\Sigma}_\pi\equiv\widetilde{\Sigma}_\pi(N)+\widetilde{\Sigma}_\pi(\Delta).$
These are shown in Fig.~\ref{fig:extrap} as a function of the varying pion mass $\mu$, the only parameter that is varied.
 Bare masses, $M_0=2.42\;{\rm GeV},\;\widetilde{M}_0= 2.06 $ GeV have been added to  $\Sigma_\pi,\widetilde{\Sigma}_\pi $ so as to reproduce the lattice data point at $\mu/{4\pi f_\pi}= 0.252 (\mu=293)$ MeV. We use $M_A=1.03 $ GeV. 
  The use of the exact expression gives an approximately  linear dependence on the pion mass, in agreement with the 
   ``surprisingly linear" results of lattice QCD simulations~\cite{hep-lat/0608010,arXiv:0806.4549}, found for values of  $\mu$ greater than about 290 MeV. The  LHP lattice data~\cite{arXiv:0806.4549}
   are shown, and these are consistent with other lattice calculations as reviewed 
    Varying the value of $M_A$ within the stated range changes the value of $\Sigma_\pi$ only for $\mu>0.5$ GeV, and by 5 \% or less.
   The low-order chiral approximation of  \eq{nt} and \eq{dt} fails badly, showing that the chiral logarithms do not dominate for the relatively large values of $\mu$  used 
   in many previous lattice QCD calculations. One could carry out the expansions of  \eq{nt} and \eq{dt} to higher order in $\mu$, but convergence requires many terms. One achieves a satisfactory description of $\Sigma(N)$ up to $\mu=0.65$ GeV by keeping terms up to order $\mu^{24}$, and of 
   $\Sigma(\Delta)$  up to $\mu=M_\Delta-M$ GeV by keeping terms up to order $\mu^{20}$.

It is worthwhile to compare our procedure with that  of some others. For example,  if one uses the heavy baryon limit to simplify  
\eq{sigma1}, evaluates the integral by taking the pion to  be on its  mass shell and regularizes the divergent 
 integral over momentum using a cutoff at a maximum momentum, one obtains results that correspond to the terms used in 
\cite{Leinweber:1999ig}.
   The relativistic procedure of \cite{Becher:1999he} avoids the use of the heavy baryon limit by  treating the nucleon
recoil terms using an expansion procedure and uses dimensional regularization. We include all of the recoil terms  and employ a cut-off procedure that is constrained by experimental data.  In chiral perturbation theory, our procedure  corresponds to keeping a  specific set  of higher-order terms  with
a fixed relation between them,  a relation fixed by experimental data.

  Our results do not include
contributions of order higher  than $1/f_\pi^2$.  These may be considered as keeping the lowest order pion cloud corrections using an expansion in powers of $\varepsilon\equiv1/(4\pi f_\pi R)^2$, where $R$ is a confinement radius~\cite{Theberge:1980ye,Jaffe:1979df,DeTar:1980rs}.
Here $R\sim \sqrt{12}/M_A$~, so $\varepsilon \approx 1/12$. Thus we expect our  results for the terms computed here to be accurate within about 10\%. This argument was mainly applied to terms involving combinations of couplings of the nucleon to a single pion, but also holds for the $n$-pion-nucleon vertex {\it e.g.} as appearing in Fig.~\ref{fig:diag}d.
These terms  enter at higher orders in  $\mu$ 
in chiral perturbation theory~\cite{WalkerLoud:2004hf}. The coupling constant $g_{\pi N}$ and the confinement sizes of the pion and the nucleon, although not explicit in chiral perturbation theory,  enter into the calculation of the diagram in terms of quarks and gluons and via the implicit dependence of $f_\pi$ and $g_{\pi N}$ on the underlying  strong coupling constant, $\alpha_S$. Therefore we expect that the terms of the chiral Lagrangian will be consistent with  the expansion in $\varepsilon$.

\begin{figure}
\includegraphics[width=9.991cm,height=8cm]{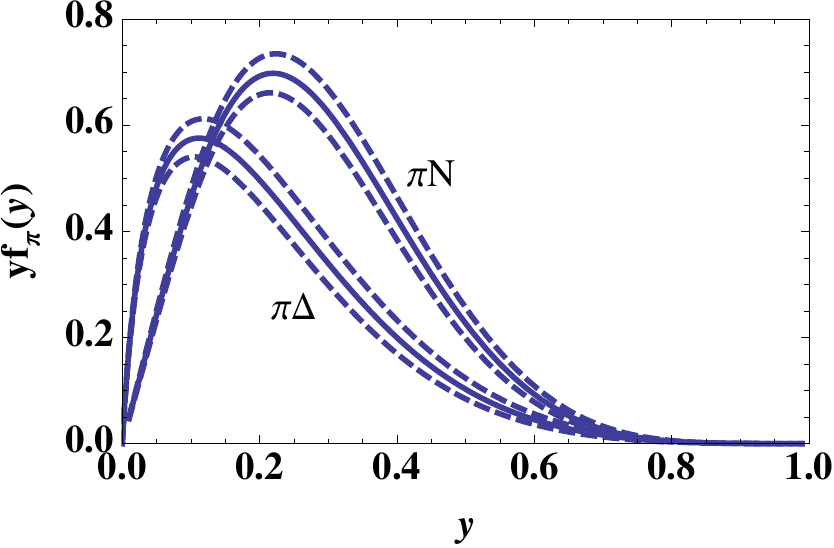}
\caption{$y f_\pi(y)$ for the intermediate $\pi N$ and $\pi\Delta $ states for  $M_A$ = 0.99,1.03,1.07  GeV} \label{fig:SFPlot}\end{figure}

To  test   our treatment of the nucleon self-energy,  we  consider the contribution to  lepton-nucleon deep inelastic scattering DIS  arising from  virtual pions. This is related to the  term $ {\cal M}_\pi$, obtained  from  Feynman rules for the diagram of Fig 1b, as
\bea {\cal M}_\pi=
2M {\partial \Sigma_\pi\over\partial \mu^2}.
\label{pion}\eea
This  expression   does not involve  a ``probability", because the square of a nucleon light-front  wave function does not appear.
Note that charge and momentum are explicitly conserved: production of a pion of momentum $k$ is accompanied by an intermediate  nucleon of momentum $p-k$. The integrations over 
$k^-,k_\perp$ are carried out explicitly, and with the definition $y=k^+/p^+$ one finds
\bea
&&{\cal M}_\pi=\int_0^1dy f_\pi(y),\;  f_\pi(y)\equiv f^N_\pi(y)+f^\Delta_\pi(y),\nn
 f^N_\pi(y)\equiv {3 g_{\pi N}^2}{\pi\over 2(2\pi)^3}\int_{y^2M^2/(1-y)}^\infty  dt\; {t\;F^2(-t)\over (t+\mu^2)^2}, \nn
f_\pi^\Delta(y)\equiv2 ({ g_{\pi N\Delta}\over 2M})^2{\pi\over   (2\pi)^3}
 { 2\over3}\int_{(y^2M^2+y(M_\Delta^2-M^2))/(1-y)}^\infty  dt\; {\;F^2(-t)\over (t+\mu^2)^2}\nn\times
\left (t+{1\over4M_\Delta^2}(M^2-M_\Delta^2+t)^2\right)
{1\over2}((M+M_\Delta)^2+t) .
\eea
The functions $f_\pi^N(y),f^\Delta_\pi(y)$ are shown in Fig.~\ref{fig:SFPlot} where one   observes that these functions  are of roughly equal importance.

The change in the quark distribution functions of the nucleon, $\delta q_i(x)$, from this effect is given by the convolution formula as
$\delta q_i(x) =\int _x^1dy f_\pi(y) q_{i}^\pi(x/y),$
with $q_{i}^\pi$ the distribution functions for quarks of flavor $i$ in the pion.
The related  contribution to the nucleon structure function $\delta F_2(x)$ is
 \bea \delta F_2(x)=
 \int_x^1 y f_\pi(y) F_{2}^{\pi}(x/y)dy,\eea 
where $F_{2}^{\pi}$ is the pion structure function~\cite{Thomas:1983fh,Ericson:1983um}. 

An integral involving the difference between the proton and neutron structure functions is particularly  interesting:
\bea \int_0^1 {dx\over x} ( F_{2}^{p}(x)-F_{2}^{n}(x))=1/3 -{2\over3}\int_0^1 dx(\bar{d}(x)-\bar{u}(x)),\eea
where the first term, obtained if the bare
 nucleon has a symmetric sea, i.e. $\bar{d}=\bar{u}$,  represents the Gottfried sum rule~\cite{Gottfried:1967kk}.  
Experiment has clearly established violation of the Gottfried sum rule, and the most precise determination of the sea asymmetry \cite{Towell:2001uq} is
\bea D\equiv\int_0^1 (\bar{d}(x)-\bar{u}(x))dx=0.118 \pm 0.012.\label{exp}\eea
Henley \& Miller \cite{Henley:1990kw} showed that
the pion cloud provides a natural explanation of the measured asymmetry.   For Fig.~\ref{fig:diag}b, the pion cloud of a proton will include $\pi^+ (u\bar d)$ and the $\pi^0$, which has equal numbers of $\bar{d}$ and $\bar{u}$. Only valence quarks of the pions are considered; the pion sea distributions are assumed to be symmetric. The probability for a $\pi^+ n$ intermediate state is 2/3, and that for a $\pi^0 p$ state is 1/3. Including also the effects of an intermediate $\Delta$ leads to 
\bea D_\pi =\int_0^1dyy ({2\over 3}f^N_\pi(y)-{1\over3}f^\Delta_\pi(y)),\eea 
with the probability of $\pi^-\Delta^{++}$ = 1/2 and that for  $\pi^+\Delta^{0}$ =1/6. Since a bare baryon is assumed to have a symmetric sea, possible contributions from Fig.~\ref{fig:diag}c do not enter.
Using 
 $M_A=1.03 $ GeV,  the nucleonic contribution is 
 0.173,  and the $\Delta$  contribution is -0.064,   so that the total is 0.109, 
 within  the experimental range of \eq{exp}. 
 
 To summarize: our light front treatment of the relevant Feynman diagrams reveals that the pion-baryon vertex function appears only between on-mass-shell baryons. This allows the vertex function to be expressed in terms of one variable,   the invariant momentum transfer $t$, and to be constrained by experimental data.
 All ambiguities regarding the theoretical input needed to evaluate effects of the pion cloud to second-order in the coupling constants for the effects of intermediate $N,\Delta$  are resolved. The uncertainty due to the neglect of higher-order terms is estimated to be about 10\%.
 Our procedure reproduces  the observed linear dependence of the nucleon mass on the pion mass found in lattice QCD calculations and the flavor asymmetry of the nucleon sea.
This work has implications for nucleon-nucleon scattering because one is instructed to use the coupling implied by 
 \eq{relate}, and also for computing pion cloud effects on the elastic electromagnetic form 
 factors of nucleons~\cite{Miller:2002ig}.

\section*{Acknowledgements} This   work has been partially supported by 
U.S. D. O. E.  Grant No. DE-FG02-97ER-41014 and NSF Grant No. 0855656. GAM thanks Adelaide University for its hospitality  when some of this work was carried out. We thank M. J. Savage for useful discussions and A. Walker-Loud for providing the LHP data points.

 \end{document}